\documentclass{article}
\usepackage[T1]{fontenc}
\usepackage[utf8]{inputenc}
\usepackage{ismir}
\usepackage{amsmath,amssymb,cite,url}
\usepackage{graphicx}
\usepackage{color}
\usepackage{booktabs}
\usepackage{multirow}
\usepackage{algorithm}
\usepackage{algorithmic}
\usepackage{bm}
\usepackage[normalem]{ulem}

\newcommand{\inlineSubsection}[2]{
  \refstepcounter{subsection} 
  \vspace{0.1cm}\noindent\textbf{\thesubsection\ #1}\label{#2}
}

\usepackage{xcolor}
\usepackage{tikz}

\definecolor{accfill}{HTML}{DAE8FC}
\definecolor{mixfill}{HTML}{F8CECC}

\definecolor{accstroke}{HTML}{6C8EBF}
\definecolor{mixstroke}{HTML}{B85450}

\DeclareRobustCommand{\mixicon}{%
  \tikz[baseline=-0.6ex]\filldraw[
    fill=mixfill, draw=mixstroke, line width=0.35pt
  ] (0,0) circle (0.9ex);%
}

\DeclareRobustCommand{\accicon}{%
  \tikz[baseline=-0.6ex]\filldraw[
    fill=accfill, draw=accstroke, line width=0.35pt
  ] (0,0) circle (0.9ex);%
}

\DeclareRobustCommand{\noiseicon}{%
  \tikz[baseline=-0.6ex,x=1ex,y=1ex]{
    \clip (0,0) circle (0.95);
    \fill[white] (-1.1,-1.1) rectangle (1.1,1.1);
    \foreach \x/\y/\r/\g in {
      -0.72/ 0.62/0.10/35,
      -0.45/ 0.18/0.09/65,
      -0.18/ 0.70/0.08/45,
       0.05/ 0.38/0.10/75,
       0.28/ 0.68/0.08/30,
       0.58/ 0.25/0.09/55,
       0.78/ 0.58/0.08/80,
      -0.62/-0.10/0.09/60,
      -0.30/-0.48/0.10/40,
       0.08/-0.18/0.08/85,
       0.38/-0.58/0.10/50,
       0.74/-0.12/0.09/35,
      -0.08/-0.78/0.09/70,
       0.52/-0.76/0.08/55
    }{
      \fill[black!\g] (\x,\y) circle (\r);
    }
    \draw[gray!65, line width=0.35pt] (0,0) circle (0.95);
  }%
}

\title{{LiveBand}: \\Live Accompaniment Generation in the Audio Domain}

\multauthor
{
\begin{tabular}{c}
Marco Pasini$^2$ \hspace{0.5cm} Javier Nistal$^1$ \hspace{0.5cm} Ben Hayes$^1$ \hspace{0.5cm} Mathias Rose Bjare$^3$ \\ Stefan Lattner$^1$ \hspace{0.5cm} György Fazekas$^2$
\end{tabular}
}
{
$^1$Sony Computer Science Laboratories, Paris, France\\
$^2$Queen Mary University of London, UK \hspace{1cm}
$^3$Johannes Kepler University, Linz, Austria\\
{\tt m.pasini@qmul.ac.uk}
}

\def\authorname{M. Pasini, J. Nistal, B. Hayes, M. R. Bjare, S. Lattner, and G. Fazekas}

\usepackage[bookmarks=false,pdfauthor={\authorname},pdfsubject={\pdfsubject},hidelinks]{hyperref}

\begin{document}

\maketitle

\begin{abstract}

We present LiveBand, a real-time system that generates high-fidelity music accompaniments to live audio input, respecting strict causal constraints. Our method trains a causal transformer generator in the continuous latent space of a pre-trained causal audio autoencoder, using adversarial sequence-level supervision from a discriminator. At each timestep, the generator receives only the causally available mix context and Gaussian noise, and predicts accompaniment latents without access to future mix frames or ground-truth target latents. Training is performed in a single parallel forward pass under causal masking, while streaming inference proceeds autoregressively with a rolling attention state. The model’s training and inference computations are matched by design, eliminating teacher forcing and the associated exposure bias. On a multi-instrument music accompaniment benchmark, LiveBand improves over prior work on objective measures of audio quality, beat alignment, and mix adherence, while enabling real-time streaming generation without lookahead into the future on consumer hardware. The demo page can be found here.\footnote{\href{https://sonycslparis.github.io/liveband-companion}{https://sonycslparis.github.io/liveband-companion}}
\end{abstract}

\section{Introduction}\label{sec:introduction}

Designing AI systems to jam--creating musical accompaniments in real time while listening to a live audio stream--is a long-standing goal at the intersection of music information retrieval, generative modelling, and human-computer interaction. Such a system would enable musicians to jam with an AI companion that responds naturally to their playing, adapting to tempo, key, dynamics, and timbre. The central challenge is streaming generation under strict causality: the model must produce each output frame at low latency and without access to future input, while maintaining musical coherence with the performer and its own past generation.

Real-time music generation has been demonstrated in unconditioned or weakly conditioned settings~\cite{livemusicmodels}, and offline systems have shown strong results for accompaniment generation~\cite{diffariff,improveddiffariff,stemgen}. However, the specific problem of \emph{live, mix-conditioned accompaniment} remains substantially more difficult. Recent work based on causal transformers over discrete neural audio tokens suggests that achieving adherence between the generated accompaniment and the input may require some lookahead into the future mix~\cite{wu2025streaming}.

In this paper, we argue that this limitation is not fundamental to the task itself, and we propose a different training objective that overcomes it. 
We present LiveBand, a real-time generative model that replaces the purely next-step predictive supervision of previous accompaniment models with sequence-level supervision using a conditional adversarial discriminator. We also avoid the use of teacher forcing during training. As a result, the model can generate accompaniments with strong mix adherence and without error accumulation, even when operating \emph{ahead} of the live input, i.e., generating a future accompaniment frame without access to the currently playing mix. This is required for real-world viability, where latency constraints need to be accounted for. In essence, we show that generating one step ahead is not a fundamental barrier, but rather a supervised learning mismatch that can be resolved with the right training paradigm.

\begin{figure*}[t]
  \centering
  \begin{minipage}[t]{0.49\textwidth}
    \centering
    \includegraphics[width=\linewidth]{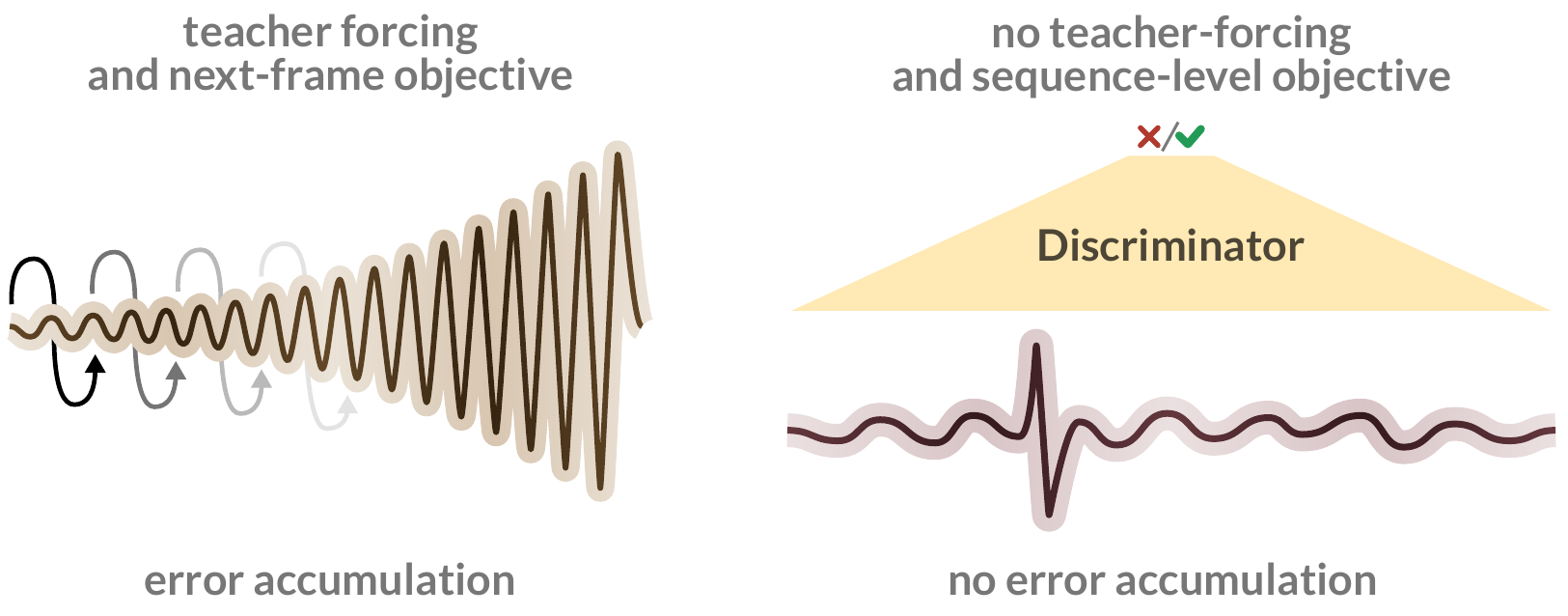}
  \end{minipage}\hfill
  \begin{minipage}[t]{0.49\textwidth}
    \centering
    \includegraphics[width=\linewidth]{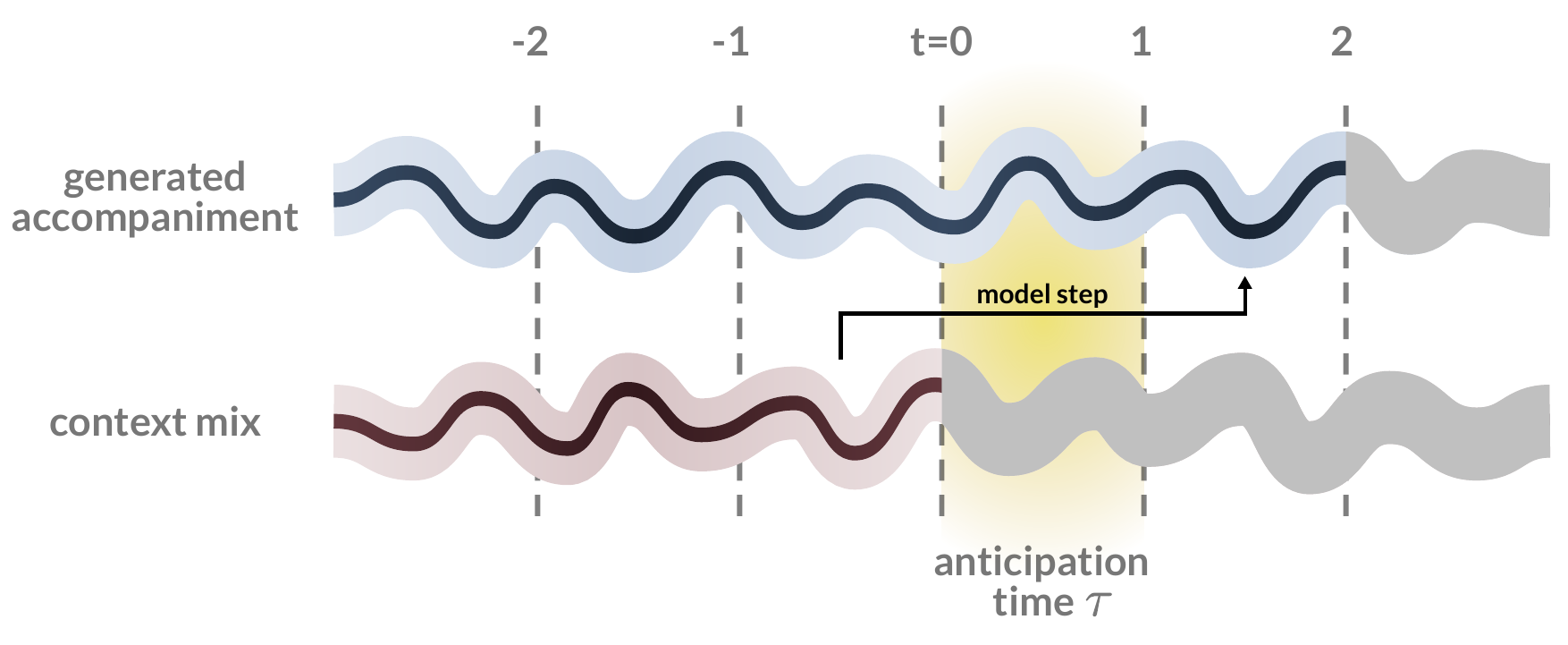}
  \end{minipage}
  \caption{Overview of the proposed streaming setting and training intuition. Sequence-level supervision directly rewards globally plausible continuations instead of step-wise target matching, and the delayed-causality setup provides a practical real-time inference budget. \textbf{Left panel}: next-step forcing can accumulate local errors over long roll-outs, while sequence-level adversarial supervision with no teacher forcing evaluates whole sequences and avoids drift. \textbf{Right panel}: at model step $t$, generation uses only causally available mix context up to $t$ (i.e, $m_t$) to produce  an accompaniment frame at time $t+1+\delta$ (i.e. $a_{t+1+\delta}$); the highlighted window denotes physical anticipation time $\tau = \delta / f_{\text{lat}}$, which provides compute headroom for real-time inference under strict causality.
  }
  \label{fig:teaser}
\end{figure*}

Our approach is motivated by two key observations. First, common training paradigms for autoregressive generation must contend with a distribution shift between training and streaming inference: teacher forcing suffers from exposure bias~\cite{arora2022exposure}, and while self-forcing can reduce this mismatch, it does so at the cost of sequential iterative rollout during training~\cite{huang2025selfforcing}. By using independent per-step noise as generator input, our model instead generates the full sequence in a single parallel causal pass during training while matching the inference-time distribution by design. Second, real-time accompaniment inherently conflicts with frame-level objectives. Because future context is unknown, even expert musicians in this setting would commit minor anticipatory errors that require continuous, on-the-fly realignment, and frame-level objectives rigidly penalize these natural corrections. To address this, we use a discriminator that evaluates entire generated sequences; this provides distribution-level supervision, ensuring the model focuses on the holistic realism and coherence of the sequence rather than independent frame accuracy. 
As a result, the model is encouraged to preserve longer-range musical structure while tolerating small local timing deviations. 
Figure~\ref{fig:teaser} illustrates this intuition.

Our contributions are threefold:
(1)~We introduce a strictly causal transformer for real-time music accompaniment, trained with adversarial sequence-level supervision and operating without lookahead. (2)~To the best of our knowledge, this is the first streaming accompaniment model to fully match training and inference conditions by design, thereby eliminating the exposure bias induced by teacher forcing without requiring sequential rollout during training. (3)~We propose an adaptive gradient penalty weight that automatically keeps the discriminator at a fixed advantage over the generator, stabilising adversarial training without manual tuning.

\section{Related Work}\label{sec:related} 
Real-time accompaniment generation builds on prior work in both automated and interactive music generation. Early systems range from rule-based and symbolic approaches~\cite{dannenberg1984scorefollowing,musictransformer,songdriver,biles1999genjam,realjam,ganmusicinteraction,anticipatorytransformer} to recent neural models operating directly on audio~\cite{musika,bassnet,bassnet2,drumgan,stemgen,diffariff,improveddiffariff}. While these latter systems show that high-quality accompaniment can be learned from acoustic context, they are typically designed for offline use, where future context is available and strict latency constraints do not apply. A separate line of work studies real-time generative systems in interactive prompt-driven settings~\cite{livemusicmodels}, but these do not address accompaniment from a live incoming mix under strict causality.

Only recently has accompaniment been formulated explicitly as a streaming latent-generation problem~\cite{wu2025streaming,realtimeaccompanimentgen}. This line of work shows that live accompaniment is feasible, but also suggests that strong mix adherence becomes difficult without some degree of future lookahead. Our work differs in two key respects: we replace next-step predictive supervision with adversarial sequence-level supervision, and we eliminate teacher forcing entirely by driving each step only with causal mix conditioning and independent noise. This matches training and inference by design. In this sense, LiveBand connects recent streaming accompaniment work with adversarial sequence modeling for audio and music~\cite{timegan,wavegan,gansynth,musika, vqcpcgan, r3gan}, while targeting the stricter real-time setting considered here.

A closely related line of work, especially in autoregressive video generation, studies \emph{drift} and long-horizon error accumulation under self-conditioned rollouts. Foundational sequence-model analyses tie this failure mode to exposure bias from teacher forcing~\cite{ranzato2016seqtraining,arora2022exposure}. Recent methods reduce this mismatch by finetuning on self-generated histories~\cite{huang2025selfforcing,selfforcing++}, by mixing autoregressive conditioning with token-wise diffusion noise schedules and noise augmentation~\cite{diffusionforcing, cam}, or by redesigning teacher-student supervision for long context and real-time distillation~\cite{contextforcing,causalforcing,liu2025rolling}. Parallel work shows that using KV-cache can also induce drift, and that attention-sink mechanisms can improve long-horizon consistency under KV eviction~\cite{streamingllm, liu2025rolling, openai2025gptoss}. A recent work~\cite{novack2026live} uses self-forcing to finetune diffusion models for interactive music generation. Compared to these approaches, our framework eliminates drift without expensive sequential training rollouts, and works end-to-end with a single objective, avoiding any finetuning or post-training.

\vspace{-0.2cm}
\section{Background}\label{sec:background}
\vspace{-0.2cm}
\inlineSubsection{Teacher/Student Forcing Drift}{sec:exposure-bias}
Autoregressive models are usually trained with \emph{teacher forcing}:
\begin{equation}
\mathcal{L}_{\text{TF}}=-\sum_t \log p_\theta(x_t\mid x_{<t}^{\star}),
\end{equation}
where $x_{<t}^{\star}$ is the ground-truth past. At inference, the same model must instead sample
\begin{equation}
\hat x_t\sim p_\theta(x_t\mid \hat x_{<t}).
\end{equation}
This mismatch is exposure bias: the model is optimized on histories from the data distribution, but deployed on histories from its own rollout distribution~\cite{ranzato2016seqtraining,arora2022exposure}. Intuitively, if an early prediction is slightly wrong, that error becomes part of the next input context; subsequent predictions are then made from an off-distribution state, which increases the chance of further errors. Repeating this feedback loop causes drift and long-horizon error accumulation. In real-time accompaniment, this drift is particularly fatal, as a slightly misaligned beat can quickly cascade into complete musical asynchrony.

\inlineSubsection{KV Cache Drift}{sec:kvcache}
Transformer models cache Key and Value activations during decoding to avoid recomputing attention at every step. This caching is essential for inference efficiency; however, the cache can only store a limited history before memory becomes prohibitive, or before the inference context length becomes higher than the one used during training (N). Rolling KV caching, keeping only the most recent N KVs, leads to sharp performance degradation when the model needs to attend to initial tokens that have been evicted from cache. \cite{streamingllm} shows how transformers naturally learn to make initial tokens act as \emph{attention sinks}, storing the majority of the attention mass, due to the causal training setup where initial tokens are visible to all subsequent positions. Using explicit attention sinks and not evicting their KV cache during rollouts dramatically limits performance degradation in the streaming setting.

\inlineSubsection{Relativistic GANs}{sec:r3ganbackground}
Generative Adversarial Networks (GANs~\cite{gan}) train a generator $G$ and discriminator $D$ in a two-player game: $G$ maps noise $z\sim p(z)$ to a sample $\hat{\mathbf{x}}=G(z)$, while $D$ assigns higher scores to real data than fake data. A common non-saturating objective is~\cite{hingegan}:
\begin{small}
\begin{align}
\mathcal{L}^{\text{hinge}}_D &=
\mathbb{E}[\max(0,1-D(\mathbf{x}))]+\mathbb{E}[\max(0,1+D(\hat{\mathbf{x}}))],\\
\mathcal{L}^{\text{hinge}}_G &=
-\mathbb{E}[D(\hat{\mathbf{x}})].
\end{align}
\end{small}
In practice, GAN training is sensitive because $G$ and $D$ are updated simultaneously and can oscillate or collapse modes if $D$ becomes poorly behaved.

We therefore adopt R3GAN~\cite{r3gan}, which combines a relativistic discriminator~\cite{jolicoeur2019relativistic} with zero-centred gradient penalties~\cite{gp}. Given a real sequence $\mathbf{x}$ and a generated sequence $\hat{\mathbf{x}}$, the discriminator and generator losses are

\begin{small}
\begin{align}
\mathcal{L}_D &=
\mathbb{E}\!\left[\mathrm{softplus}\!\left(
D(\hat{\mathbf{x}})-D(\mathbf{x})\right)\right],
\label{eq:dloss}\\
\mathcal{L}_G &=
\mathbb{E}\!\left[\mathrm{softplus}\!\left(
D(\mathbf{x})-D(\hat{\mathbf{x}})\right)\right],
\label{eq:gloss}\\
\mathcal{R} &= \frac{1}{2}\left(
\left\|\nabla_{\mathbf{x}} D(\mathbf{x})\right\|^2 +
\left\|\nabla_{\hat{\mathbf{x}}} D(\hat{\mathbf{x}})\right\|^2
\right),
\label{eq:gp}
\end{align}
\end{small}
where $\mathcal{R}$ is the combined R1+R2 gradient penalty~\cite{gp,
karras_analyzing_2020}. $D$ learns to rank real samples above generated ones \emph{pairwise}, so generator updates are driven by relative realism instead of absolute logit calibration. This empirically provides more informative gradients and avoids mode collapse. The R1+R2 penalties regularize $\|\nabla_x D(x)\|$ on both real and generated samples, preventing overly sharp decision surfaces, reducing oscillations, and improving training stability.


\begin{figure}[ht]
  \centering
    \includegraphics[width=\linewidth]{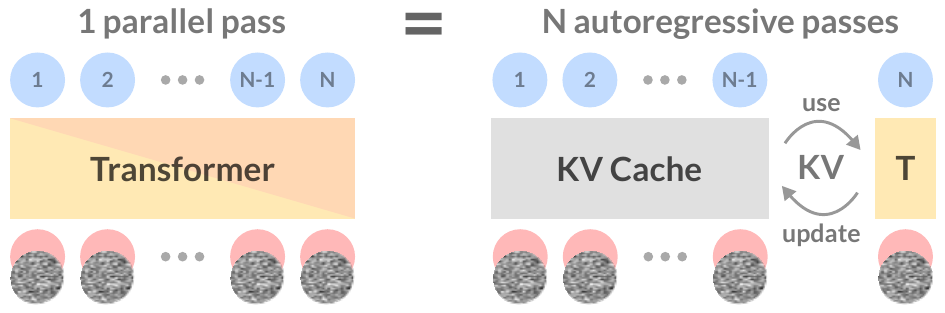}
  \caption{
  \mbox{\mixicon~mix},
  \mbox{\accicon~acc.},
  \mbox{\noiseicon~noise}. Training-inference equivalence of the generator. A full sequence can be produced in one parallel causal pass during training, while inference executes $N$ autoregressive steps with a KV cache; under the same causal mask and inputs, both procedures are computationally equivalent.}
  \label{fig:equivalence}
\end{figure}

\vspace{-0.4cm}
\section{LiveBand}\label{sec:model}
\vspace{-0.2cm}
\inlineSubsection{Problem formulation}{sec:problem}
We consider the task of real-time accompaniment generation from a live input mix. Let $\mathbf{m} = (m_1, m_2, \ldots)$ denote the sequence of mix latent frames and $\mathbf{a} = (a_1, a_2, \ldots)$ the sequence of accompaniment latent frames to be generated. At streaming step $t$, the model has access to the causally available mix history $\mathbf{m}_{\leq t} = (m_1, \ldots, m_t)$ and predicts a future accompaniment frame $a_{t+1+\delta}$, where $\delta \geq 0$ denotes the \emph{anticipation} in latent frames. Thus, $\delta$ measures how far ahead of the observed mix the accompaniment is generated, and the model never has access to future mix information.

Both mix and accompaniment are represented in the continuous latent space of a pre-trained causal audio autoencoder (Section~\ref{sec:codec}), so the task is formulated directly in latent space. The corresponding anticipation in seconds depends on the latent frame rate $f_{\text{lat}}$. We therefore distinguish between the discrete anticipation $\delta$ (in latent frames) and its temporal equivalent $\tau = \delta / f_{\text{lat}}$ (in seconds). We use $\delta$ in formal definitions and $\tau$ when discussing latency and reporting operating points in tables. Note that $\tau > 0$ is strictly required in real-world deployment, since it provides a non-zero time budget needed to observe the input mix, compute the next latent prediction, and decode it to audio with the neural decoder before playback. The right panel of Figure~\ref{fig:teaser} illustrates this operating regime and the temporal offset between observed mix context and generated accompaniment, where inference is performed.

\inlineSubsection{Causal Audio Autoencoder ($AE$)}{sec:codec} We use a pre-trained causal audio autoencoder obtained by retraining a causal variant of CoDiCodec~\cite{codicodec}. The model encodes stereo waveforms at 44.1\,kHz into $\sim10$\,Hz continuous latent sequences with dimension $d_{\text{latent}} = 128$ and a temporal downsampling ratio of $4096\times$. In contrast to the original CoDiCodec formulation, we use only the continuous latent pathway, expand the bottleneck dimension, and modify both encoder and decoder to operate strictly causally, so that neither stage has access to future audio context. We use the same training data and training details as CoDiCodec. The autoencoder is trained separately and then kept fixed during training of the generative model.
Before being passed to the generative model, latent representations are normalised to zero mean and unit variance per channel using statistics computed over the training set.

\inlineSubsection{Generator ($G$)}{sec:generator}
The generator is a causal transformer operating on latent frames. At each timestep $t$, it receives two external inputs: an independent Gaussian noise vector $\mathbf{z}_t$ and the conditioning mix latent frames $\mathbf{m}_{t}$. These are concatenated and linearly projected to the transformer hidden dimension. The generator is thus conditioned only on causally available mix information and per-step noise.

The backbone consists of pre-norm transformer blocks. Each block comprises adaptive layer normalization $\mathbf{c}$~\cite{adain}, a causal multi-head self-attention layer with query--key normalization~\cite{henry2020qknorm} and Rotary Position Embeddings~\cite{su2021rope}, and a SwiGLU feed-forward network~\cite{shazeer2020glu}. We implement causal masking with scalar attention sink ~\cite{streamingllm,openai2025gptoss} jointly using FlexAttention~\cite{he2024flexattention}. After the final block, the hidden representation is passed through a final adaptive layer normalization conditioned on $\mathbf{c}$ and then projected back to the autoencoder's latent space.
The generator is further conditioned on a learned instrument-class embedding, together with a null class used when class conditioning is dropped. Concrete architectural hyperparameters are reported in the training details in Sec.~\ref{sec:training}.


\inlineSubsection{Training-Inference Equivalence}{sec:parallel}
Autoregressive models suffer from exposure bias when training on ground-truth histories but inference runs on the model's own predictions (see Sec.~\ref{sec:exposure-bias}). Our generator avoids this mismatch by construction: at streaming step $t$, it receives a noise vector $\mathbf{z}_t$ together with the causally available mix $\mathbf{m}_{\leq t}$, and predicts the future accompaniment frame $a_{t+1+\delta}$ (see Sec.~\ref{sec:problem}). Previous target latents are never fed back as inputs; instead, temporal dependencies are carried internally by self-attention over the hidden-state sequence. We also use a scalar sink to limit KV cache-based drift during long-form streaming (see Sec.~\ref{sec:kvcache}).
This yields a key property: parallel masked computation during training is exactly equivalent to step-wise autoregressive computation at inference. Under a causal attention mask, the hidden state at position $t$ in the parallel pass is identical to that obtained by processing positions $1,\dots,t$ sequentially with KV cache, since no information can flow from future positions. As a result, the full training sequence can be generated in a single forward pass, while inference proceeds one frame at a time as new mix input conditioning arrives. Because the generator receives the same input in both cases (independent noise and causal mix context), training and inference are matched by design. Unlike forcing-based approaches~\cite{huang2025selfforcing}, this does not require expensive sequential rollout during training, and due to our adversarial objective each frame is also generated in a single step at inference time (see Fig.~\ref{fig:equivalence}).

\inlineSubsection{Discriminator ($D$)}{sec:discriminator}
The discriminator is a non-causal 1D convolutional network that scores the realism of a complete accompaniment sequence conditioned on the corresponding mix. We use non-causal convolutions since $D$ is only used at training time, and may provide more informative signal to $G$ by having access to the full accompaniment and mix contexts. We also do not notice any training instability with this asymmetry in causality between $D$ and $G$. The input is the concatenation of the accompaniment latents $\mathbf{a} \in \mathbb{R}^{T \times d_{\text{latent}}}$, either real or generated, and the mix latents $\mathbf{m} \in \mathbb{R}^{T \times d_{\text{latent}}}$, yielding a joint sequence representation in $\mathbb{R}^{T \times 2d_{\text{latent}}}$. This representation is processed by a hierarchy of convolutional residual blocks with progressive temporal downsampling, followed by a final linear layer that outputs a single scalar logit.
Instrument-class conditioning is injected throughout the discriminator via learned scaling in the residual blocks, similar to AdaLN~\cite{adain} used in $G$ but with no normalization. Following R3GAN~\cite{r3gan}, the discriminator uses no normalization layers.

\inlineSubsection{Adaptive Gradient Penalty (AdaGP)}{sec:adaptive_gp}
Adversarial training is sensitive to the balance between generator and discriminator, which in our setting is controlled mainly by the gradient-penalty weight $\lambda$ in the discriminator objective $\mathcal{L}_D + \lambda \mathcal{R}$. If $\lambda$ is too small, the discriminator becomes overly strong and yields unstable gradients; if too large, it becomes too weak to guide the generator effectively. The optimal gradient penalty weights can vary wildly depending on the training setting~\cite{r3gan}. To reduce this tuning burden, we adapt $\lambda$ online to maintain a target $D$ advantage over $G$. Defining $\mathrm{adv}=\mathbb{E}[D(\mathbf{x})]-\mathbb{E}[D(\hat{\mathbf{x}})]$, we update $\lambda \leftarrow \max\!\bigl(0,\lambda+\eta\,\mathrm{sign}(\mathrm{adv}-a^*)\bigr)$, where $\eta$ is a fixed step size and $a^*$ is the target advantage. When the discriminator becomes too strong ($\mathrm{adv}>a^*$), $\lambda$ is increased to strengthen regularization; when it becomes too weak ($\mathrm{adv}<a^*$), $\lambda$ is decreased. In this way, AdaGP keeps the discriminator in a regime where it remains informative without dominating training.

\inlineSubsection{Implementation Details}{sec:training}
We train LiveBand with the R3GAN framework (Sec.~\ref{sec:r3ganbackground}) in the frozen latent space of the causal autoencoder (Sec.~\ref{sec:codec}). Training uses latent crops of length $T=128$ frames ($\sim12$\,s), and our main model operates at $\tau \approx 0.093$\,s anticipation, meaning that our theoretical inference budget for generator and audio decoder is also $\tau \approx 0.093$\,s (Fig.~\ref{fig:teaser}). The generator uses per-step Gaussian noise of dimension $d_z=32$, model dimension $d_{\text{model}}=1024$, $L=12$ transformer blocks, and $H=8$ attention heads of dimension $d_h=128$. The discriminator uses fixed width $C=1536$ and five temporal resolution levels, and each convolutional block is a ConvNext-style block~\cite{convnext} as in~\cite{r3gan}, using grouped convolutions with $64$ groups. Both networks are conditioned on 18 MIDI instrument categories plus a null class, with conditioning dropout probabilities $p_{\text{mix}}=0.1$ and $p_{\text{cls}}=0.1$.
We optimize both networks with Adam~\cite{adam} using $\beta_1=0.5$, $\beta_2=0.9$, no weight decay, and fixed learning rate $10^{-4}$ with 1k warm-up steps. AdaGP (Sec.~\ref{sec:adaptive_gp}) uses target advantage $a^*=1.0$ and adaptation step size $\eta=0.1$. R1/R2 penalty is applied lazily every $k=8$ steps, alternating between real and generated samples and scaling the penalty by $k$ on update steps~\cite{karras_analyzing_2020}. Training uses mixed precision (\texttt{bfloat16}) with \texttt{torch.compile}, batch size 128, and 750k iterations on a single RTX~3090 GPU ($\sim1$ week of training time). Both generator and discriminator have  $\sim150$\,M parameters.

\vspace{-0.4cm}
\section{Experiments}\label{sec:experiments}
\begin{table*}[t]
\centering
\begin{tabular}{lccccccc}
\toprule
\textbf{Model} & \textbf{$\tau\,$[s]} &
\textbf{FAD$_{\text{vgg}}\downarrow$} &
\textbf{FAD$_{\text{clap}}\downarrow$} &
\textbf{BA$_{\text{F1}}\uparrow$} &
\textbf{COC$_{\text{full}}\uparrow$} &
\textbf{COC$_{\text{harm}}\uparrow$} &
\textbf{COC$_{\text{perc}}\uparrow$} \\
\midrule

\textit{Ground truth} & -- 
& 1.22 / +0.02
& 0.08 / -0.01
& 0.60 / +0.01
& 65.18 / +0.36
& 66.32 / +0.40
& 68.67 / +0.41 \\
\specialrule{0.01em}{0pt}{2pt}

\multirow{2}{*}{SMG~\cite{wu2025streaming}} & 0 
& 2.81 / +1.37 
& 0.30 / +0.26 
& 0.30 / -0.05 
& 59.74 / -0.11 
& 61.14 / -0.06 
& 64.06 / +0.03 \\

& 1 
& 2.67 / +1.67 
& 0.29 / +0.28 
& 0.19 / -0.05 
& 54.37 / +1.48 
& 56.06 / +1.42 
& 58.89 / +1.42 \\
\specialrule{0.01em}{0pt}{2pt}

\multirow{3}{*}{LiveBand} & 0 
& 1.55 / -0.12 
& 0.31 / -0.04 
& \textbf{0.65} / \textbf{+0.02} 
& \textbf{65.22} / \textbf{+0.43} 
& \textbf{66.40} / \textbf{+0.41} 
& \textbf{68.74} / \textbf{+0.49} \\

& 0.1 
& \textbf{1.39} / \textbf{-0.08} 
& \textbf{0.31} / \textbf{-0.05} 
& \textbf{0.64} / \textbf{+0.03} 
& 65.11 / +0.35 
& 66.30 / +0.37 
& 68.67 / +0.39 \\

& 1 
& 1.68 / -0.12
& 0.32 / -0.05
& 0.60 / -0.01 
& 64.30 / +0.92
& 65.38 / +1.02 
& 67.94 / +0.95 \\
\specialrule{0.01em}{0pt}{2pt}

LiveBand$_{\text{bid}}$ & -- 
& 1.30 / ---
& 0.27 / ---
& 0.64 / ---
& 65.63 / ---
& 66.69 / ---
& 69.08 / --- \\

LiveBand$_{\text{int}}$ & 0.1 
& 1.38 / -0.04
& 0.29 / -0.04
& 0.63 / +0.03
& 64.48 / +0.31
& 65.96 / +0.33
& 68.30 / +0.35 \\
\bottomrule
\end{tabular}
\caption{Main benchmark entries are reported as \textit{first 10\,s / $\Delta_{20s}$}, where $\Delta_{20s} = s_{10\text{--}20\,\mathrm{s}} - s_{0\text{--}10\,\mathrm{s}}$. For FAD, negative drift is favorable; for Beat Alignment and COCOLA, positive drift is favorable. Drift is not reported for non-streaming models. Values are rounded to two decimals. We \textbf{highlight} the best metrics achieved in $s_{10\text{--}20\,\mathrm{s}}$ between LiveBand and SMG, excluding reference models.}
\label{tab:main_benchmark}
\end{table*}
\vspace{-0.3cm}
\inlineSubsection{Dataset}{sec:dataset} 
Unless otherwise stated, all models are trained and evaluated on the official Slakh2100 train/test split~\cite{slakh2100}. We form each training example by selecting one stem as the target accompaniment, randomly choosing a subset of $[1,\ldots, N-1]$ remaining stems, and summing them to create the conditioning mix~\cite{wu2025streaming,diffariff}. Audio is encoded with the causal autoencoder into continuous latents at ${\sim}10$\,Hz, and training uses crops of $T=128$ frames ($\sim12$\,s). We also train one CLAP-conditioned variant on an internal corpus of approximately 20k non-synthetic multitrack stereo recordings, but evaluate it on the Slakh2100 test set for comparability (LiveBand$_{\text{int}}$).

\inlineSubsection{Evaluation}{sec:eval_protocol}
For streaming-capable models, we generate 20\,s accompaniments and evaluate each output on two non-overlapping 10\,s segments, yielding a short-horizon quality estimate and a drift measure $\Delta s = s_{10\text{--}20\,\mathrm{s}} - s_{0\text{--}10\,\mathrm{s}}$. For non-streaming models, only the first 10\,s segment is reported. We also conduct a subjective listening test comparing baselines on audio quality of the generated accompaniments, and their adherence to the mix.

\inlineSubsection{Metrics}{sec:metrics}
We report Fréchet Audio Distance in the VGGish and LAION-CLAP embedding spaces (FAD$_{\text{vgg/CLAP}}$, $\downarrow$)~\cite{kilgour_frechet_2019,vggish,laionclap}, Beat Alignment F1 (BA$_{\text{F1}}$, $\uparrow$) using BeatThis and Madmom~\cite{bock_madmom_2016, beatthis}, and COCOLA (CC, $\uparrow$)~\cite{cocola}, for which we use the full, harmonic, and percussive variants. COCOLA is a contrastively-trained model measuring the adherence between mix-accompaniment pairs.

\inlineSubsection{Baselines}{sec:baselines}
We compare against StreamMusicGen (SMG)~\cite{wu2025streaming}, an autoregressive model trained on next-token prediction over RVQ codes from a causal DAC-style codec~\cite{dac}. We evaluate causal LiveBand$_\tau$ at $\tau \in \{0,0.1,1\}$\,s, corresponding to perfectly synchronous, realistic real-time use, and strongly anticipatory operating regime. We also include a sink variant (LiveBand$^{\text{sink}}_{\tau=0.1}$), a bidirectional upper bound (LiveBand$_{\text{bid}}$), and the internal-data CLAP-conditioned variant (LiveBand$^{\text{int}}_{\tau=0.1}$).

\inlineSubsection{Ablations}{sec:tasks}
We conduct five experiments: (1) a sink ablation comparing LiveBand$_{\tau=0.1}$ and LiveBand$_{\text{ sink}}$; (2) a GP ablation comparing fixed gradient-penalty weights $w \in \{1,10\}$ against AdaGP with $a^*=1$; (3) the main benchmark against SMG~\cite{wu2025streaming} across anticipation settings, including LiveBand$_{\text{bid}}$ and LiveBand$_{\text{int}}$; (4) a pairwise listening test comparing LiveBand$_{\tau=0.1}$ against SMG$_{\tau=0}$; and (5) an end-to-end latency benchmark for LiveBand$_{\tau=0.1}$ on a consumer RTX~3090, with and without compilation. Ablation models are trained for 100k iterations and use a discriminator width of $C=1024$.

\vspace{-0.2cm}
\section{Results}\label{sec:results}
\vspace{-0.1cm}
We provide audio examples at this link\footnote{\tiny\href{https://sonycslparis.github.io/liveband-companion}{https://sonycslparis.github.io/liveband-companion}}.

\begin{table}[t]
\centering
\resizebox{\columnwidth}{!}{
\begin{tabular}{lcccccc}
\toprule
\textbf{Model} & $\Delta$FAD$_{\text{vgg}}$ & $\Delta$FAD$_{\text{clap}}$ & $\Delta$Beat & $\Delta$COC$_{\text{full}}$ & $\Delta$COC$_{\text{harm}}$ & $\Delta$COC$_{\text{perc}}$ \\
\midrule
w/o sink & -0.02 & -0.06 & +0.01 & +0.29 & +0.27 & +0.34 \\
sink & -0.02 & -0.05 & +0.02 & +0.31 & +0.39 & +0.36 \\
\bottomrule
\end{tabular}}
\caption{Sink ablation reporting drift over 20\,s streaming generations at 0.1\,s anticipation. Both models are trained for $100$k iterations with discriminator width $C=1024$.}
\label{tab:sink_ablation}
\end{table}

\inlineSubsection{Sink vs. No-Sink}{}
Table~\ref{tab:sink_ablation} reports the sink ablation, isolating long-form drift at the main real-time operating point $\tau=0.1$\,s. Surprisingly, both variants remain stable over 20\,s, with favourable drift (improvement through time) across all reported metrics. The sink-enabled model is consistently but only slightly better on most coherence-oriented measures.
One possible interpretation is that adversarial training, together with a convolutional discriminator focused on local temporal realism, already biases the generator toward robust attention patterns with no over-concentration of attention mass, reducing practical sensitivity to KV-cache-based drifting.

\begin{table}[t]
\centering
\resizebox{\columnwidth}{!}{
\begin{tabular}{lcccccc}
\toprule
\textbf{Setup} & FAD$_{\text{vgg}}$$\downarrow$ & FAD$_{\text{clap}}$$\downarrow$ & Beat$\uparrow$ & COC$_{\text{full}}$$\uparrow$ & COC$_{\text{harm}}$$\uparrow$ & COC$_{\text{perc}}$$\uparrow$ \\
\midrule
Fixed GP ($w=1$) & 1.93 & 0.32 & 0.58 & 64.54 & 65.69 & 68.08 \\
Fixed GP ($w=10$) & 1.74 & 0.31 & 0.61 & 65.19 & 66.34 & 68.72 \\
AdaGP ($a^*=1$) & 1.82 & 0.32 & 0.60 & 65.09 & 66.23 & 68.59 \\
\bottomrule
\end{tabular}}
\caption{AdaGP ablation (first 10\,s metrics). Models trained for 100k iterations with width $C=1024$.}
\label{tab:adagp_main}
\end{table}

\inlineSubsection{Adaptive vs. Fixed GP}{}
Table~\ref{tab:adagp_main} reports the AdaGP ablation. AdaGP ($a^*=1$) clearly outperforms the under-regularized fixed-GP setting ($w=1$), while matching the best manually tuned fixed-weight setting ($w=10$). We view AdaGP as a convenience mechanism rather than a source of absolute gains. Instead of manually searching for the best-performing gradient-penalty weight, AdaGP adapts the regularization online and reliably steers training toward a well-performing discriminator regime with fixed advantage. We note that in our empirical observations the choice of the target advantage $a^*$ is not nearly as sensitive to the training setting as the GP weight, and $a^*=1$ reliably reaches the performance of manually-tuned GP weights across different model architectures and training settings.

\inlineSubsection{Main Benchmark}{}
Table~\ref{tab:main_benchmark} reports both short-horizon quality (first 10\,s) and drift over the following 10\,s. As expected, the bidirectional upper bound LiveBand$_{\text{bid}}$ achieves the strongest overall results. More importantly, among strictly causal models, LiveBand consistently outperforms StreamMusicGen (SMG) across all metrics and anticipation settings. At $\tau=(0, 0.1)$, LiveBand already matches the bidirectional upper bound in Beat Alignment and COCOLA. This suggests that the proposed causal model retains most of the sequence-level coherence of the offline variant despite operating under strict streaming constraints.
Surprisingly, this advantage stands even under strong anticipation. At $\tau=1$\,s, i.e., when the model predicts the accompaniment 1 second ahead of the latest observed mix frame, LiveBand still outperforms SMG at the strictly synchronous setting $\tau=0$ across all metrics: mix-adherent accompaniment remains possible even when generation must occur meaningfully ahead of the incoming mix.

Regarding drift, SMG results in metrics that worsen over time. LiveBand exhibits near-zero or favorable drift across all causal settings and most metrics. This makes intuitive sense: even for a human performer, performance would increase when more context information about the mix is present. The key result is that LiveBand does not accumulate local errors and is able to realign and refine the accompaniment, contrary to teacher-forced models.
Finally, several LiveBand configurations match or slightly exceed the ground-truth reference on Beat Alignment and COCOLA. We interpret this cautiously: these metrics reward synchronization and accompaniment-mix coherence, and the adversarial objective is mode-seeking at the sequence level. The generated accompaniment may therefore become more tightly aligned with the observed mix than the average ground-truth target stem. Also, the FAD metrics clearly capture a substantial gap in audio quality compared to ground truth audio. Overall, these results show that sequence-level supervision and a teacher-forcing-free generator make mix-adherent streaming accompaniment possible without future lookahead.

\inlineSubsection{Subjective Listening Test.}{}
We conduct a listening study comparing LiveBand$_{\tau=0.1}$ against SMG$_{\tau=0}$~\cite{wu2025streaming}, ground truth, and a low-quality anchor. Participants complete two 10-item tests: an audio-quality test on isolated stems, rated for overall quality and temporal consistency, and a mix-adherence test, rated for fit to the accompaniment mix and temporal consistency of this fit. Low anchors are generated differently for each test: in the quality test, by adding increasing random latent noise to encoded real stems; in the adherence test, by pairing stems with mismatching mixes from random validation-set songs, yielding poor but not necessarily time-varying adherence. Ratings use a 5-point Likert scale. We collected 30 participant files, of which 19 correspond to completed sessions. Since comparisons are within-subject, we use Friedman tests followed by pairwise Wilcoxon signed-rank tests with Holm--Bonferroni correction.

Table~\ref{tab:listening_test} shows that LiveBand is consistently preferred over SMG, scoring significantly higher across all four rated dimensions ($p_{\mathrm{Holm}}<0.006$): quality (2.6 vs.\ 1.9), quality consistency (3.0 vs.\ 2.2), mix adherence (3.4 vs.\ 2.2), and adherence consistency (3.1 vs.\ 2.3). Ground truth remains significantly above LiveBand in all dimensions ($p_{\mathrm{Holm}}\leq0.002$). SMG is not significantly different from the low-quality anchor on adherence ($p_{\mathrm{Holm}}=0.138$), whereas LiveBand is significantly above it ($p_{\mathrm{Holm}}=0.002$). Overall, the results support the objective evaluation: LiveBand is perceived as higher quality, more temporally stable, and better aligned with the input mix than SMG.


\begin{table}[t]
\centering
\resizebox{\columnwidth}{!}{
\begin{tabular}{lcccc}
\toprule
 & \multicolumn{2}{c}{\textbf{Quality} $\uparrow$} & \multicolumn{2}{c}{\textbf{Adherence} $\uparrow$} \\
\cmidrule(lr){2-3} \cmidrule(lr){4-5}
 & \textbf{Main} & \textbf{Cons.} & \textbf{Main} & \textbf{Cons.} \\
\midrule
Ground truth & 3.9 & 4.3 & 4.1 & 4.4 \\
Low Anchor & 1.4 & 1.6 & 2.2 & 3.4 \\
SMG~\cite{wu2025streaming} & 1.9 & 2.2 & 2.2 & 2.3 \\
LiveBand$_{\tau=0.1}$ & \textbf{2.6} & \textbf{3.0} & \textbf{3.4} & \textbf{3.1} \\
\bottomrule
\end{tabular}
}
\caption{Subjective results for perceived audio quality, mix adherence, and their corresponding temporal consistency.}
\label{tab:listening_test}
\end{table}

\inlineSubsection{Inference Speed on Consumer Hardware}{}
Table~\ref{tab:latency} reports per-step streaming latency for LiveBand on a single RTX 3090 GPU using \texttt{bfloat16} and batch size $1$. All measurements are averaged across $128$ streaming steps. The effective frame budget for $\tau=0.1\,$s is exactly $92.88$\,ms, corresponding to one latent frame ($4096$ audio samples at $44.1$\,kHz).
In eager mode, end-to-end generation plus decoding already remains within this real-time budget. With \texttt{torch.compile}, latency is substantially reduced.
These measurements confirm that LiveBand is deployable in real time on consumer hardware using the $\tau=0.1\,$s configuration.

\begin{table}[t]
\centering
\begin{tabular}{lccccc}
\toprule
\textbf{Mode} & \textbf{Gen.} & \textbf{Dec.} & \textbf{Mean} & \textbf{Median} & \textbf{RTF} \\
\midrule
Eager & 29.4 & 54.1 & 83.5 & 80.7 & 1.1$\times$ \\
Compiled & 25.5 & 17.9 & 43.6 & 29.8 & 2.1$\times$ \\
\bottomrule
\end{tabular}
\caption{Streaming latency, measured in \emph{ms}, and Real-Time Factor (RTF) for LiveBand$_{\tau=0.1}$ on RTX 3090. Values are computed from 128 measured steps.}
\label{tab:latency}
\end{table}

\vspace{-0.3cm}
\section{Conclusion}\label{sec:conclusion}
\vspace{-0.1cm}
We introduced LiveBand, a real-time system for live music accompaniment that operates under strict causal and latency constraints. By pairing a causal transformer with sequence-level adversarial supervision, we eliminate teacher forcing and the associated exposure bias. This fully aligns training with streaming inference, allowing the model to anticipate and adapt to live input without requiring future lookahead or accumulating error.
Empirically, LiveBand outperforms prior causal baselines in musical adherence and coherence while running in real-time on consumer hardware. However, limitations remain. Most notably, the generation's audio quality leaves room for improvement. Future work will focus on developing higher-fidelity causal audio autoencoders and scaling the system. Overall, LiveBand establishes that sequence-level adversarial training is a highly effective foundation for building robust, interactive, real-time generative models.

\section{Ethics Statement}
This work is intended for creative and artistic applications.  Automated accompaniment generation systems raise questions around authorship, the impact on professional musicians, and the potential for misuse in generating deceptive content.  We encourage the community to develop appropriate guidelines as these technologies mature.

\bibliography{ISMIRtemplate}

\end{document}